\documentstyle[12pt]{article}

\title{ 
KPP Front Speeds in Random Shears and the Parabolic Anderson Problem}
\author{
Jack Xin\\Department of Mathematics \& TICAM \\
University of Texas at Austin\\
 Austin, TX 78712, USA. \\\hspace{.1 in}\\
}
\date{}
\newcommand{\nit}{\noindent}
\newcommand{\no}{\nonumber}
\newcommand{\be}{\begin{equation}}
\newcommand{\ee}{\end{equation}}
\newcommand{\br}{\begin{eqnarray}}
\newcommand{\er}{\end{eqnarray}}

\newcommand{\dta}{\mbox{$\delta$}}

\newcommand{\lam}{\mbox{$\lambda$}}

\begin{document}
\baselineskip=18pt
\maketitle
\vspace{0.5 in}

\begin{abstract}
We study the asymptotics of front speeds of the reaction-diffusion
equations with Kolmogorov-Petrovsky-Piskunov (KPP) nonlinearity and 
zero mean stationary ergodic  
Gaussian shear advection on the entire plane. By 
exploiting connections of KPP front speeds with the almost sure 
Lyapunov exponents of the parabolic Anderson problem, and with 
the homogenized Hamiltonians of Hamilton-Jacobi equations, we show that 
front speeds enhancement is quadratic in the small 
root mean square (rms) amplitudes 
of white in time zero mean Gaussian shears, and it grows at the order of 
the large rms amplitudes. However, front speeds diverge logarithmically
if the shears are time independent zero mean stationary 
ergodic Gaussian fields.  

\end{abstract}
\vspace{.1 in}

{\bf \hspace{.1 in} }

\thispagestyle{empty}
\newpage
\setcounter{page}{1}
\setcounter{section}{0}
\section{Introduction}
Front propagation in random media appears in many scientific areas, 
and analysis of prototype models has been an efficient way to 
improve our understanding of fronts with complex structures \cite{Xin1}.
We shall consider reaction-diffusion (R-D) fronts moving inside a two 
dimensional random shear field  
modeled by:
\be
u_{t}={1\over 2} \bigtriangleup_{y} u + \sigma \vec{\xi} \cdot \nabla u + f(u),
\; \;\;\; y=(y_1,y_2) \in R^2, 
\label{J1} 
\ee
where $\vec{\xi}=(0,\xi(y_{1},t)))$, $\xi$ 
a mean zero stationary ergodic Gaussian field, 
$\sigma > 0$ a parameter measuring the shear strength (root mean square 
amplitude 
for short), 
$\bigtriangleup_{y} = \partial^{2}_{y_{1}y_{1}} +
\partial^{2}_{y_{2}y_{2}}$, $f(u)=u(1-u)$, known as the KPP nonlinearity.  
We are interested in fronts along the $y_{2}$
direction, and how the multi-scales in the 
shear field $\vec{\xi}$ influence the speed of propagation, a quantity at large scale 
(large time).  

If the shear $\vec{\xi}$ is deterministic, e.g. periodic in space (and time), 
the asymptotic behavior 
of front speed is known for many forms of reaction nonlinearity $f(u)$. 
For small 
$\sigma $ and zero mean $\xi $:
\be
c_{*} = c_0 ( 1 + \alpha  \sigma^2 + h.o.t ), \label{J2}
\ee
where $\alpha $ is a positive constant depending only on $\xi$, 
$c_0$ is the pure R-D front speed. The constant 
$\alpha $ is universal among bistable nonlinearity 
($f(u)=u(1-u)(u -\mu )$, $\mu \in (0,1/2)$), ignition combustion nonlinearity 
($f(u)=0, u \in (0,\theta)$, $\theta \in (0,1)$, ignition level,  
$f(u) >0$, $u \in (\theta,1)$), 
and others ($f(u)=u^m(1-u)$, $m \geq 2$), 
except KPP ($f(u)=u(1-u)$) where $\alpha$ takes 
another positive value. See calculations using perturbation and 
variational methods \cite{PX}, \cite{HPS}, in the spatially periodic case. 
The robustness of (\ref{J2}) in the spatially-temporally periodic case 
is studied recently in \cite{NX}. Except for the KPP nonlinearity, 
where linearization is possible to obtain front speeds, one needs to 
work with an exact traveling wave solution and the resulting equation it 
satisfies in order to find the constant $\alpha $.  
 
For large $\sigma $, on the other hand:
\be
c_{*} = O(\sigma ), \label{J3}
\ee
if $f(u) \geq 0$, see \cite{B1}, \cite{Const1}, \cite{KR}, 
\cite{H}, \cite{Const2}, 
and existence of limit $c_{*}/\sigma $ holds for the KPP \cite{B2}.   
The asymptotics (\ref{J3}) also hold in the spatially-temporally 
periodic shears \cite{NX}, and is numerically observed 
even for bistable nonlinearity \cite{NX}.

A natural question is whether these results remain valid for fronts 
in random shears. We approach this issue  
here for the KPP nonlinearity and in the entire space $R^2$.
We show that the KPP front speed formula is 
connected with the almost sure Lyapunov exponent of the so called 
parabolic Anderson problem, namely the exponential growth rate of the semigroup 
$e^{t L}$ with $L\cdot = \Delta \cdot + V(x,t)\cdot $, $V(x,t)$ a Gaussian 
process with mean zero. There is a rich literature on almost sure 
Lyapunov exponent \cite{CM1}, \cite{CM2}, \cite{CM3}. 
The logarithm of the linear parabolic equation 
satisfies the stochastic viscous KPZ equation in surface growth  \cite{Surf}
(or a quadratic Hamilton-Jacobi 
equation with random potential and viscosity), and the growth rate is 
also the effective Hamiltonian in the large space large time (homogenization) 
limit. Combining these two connections, we are able to simplify the 
front speed calculations. The main finding is: if $\xi=\xi(y_{1},t)$, white in 
time, mean zero stationary ergodic Gaussian process, then almost surely 
front speed in random shears is a deterministic constant $c_{*}$ with 
asymptotics:
\be
c^{*} = c_0 ( 1 + {1\over 2} \Gamma (0) \sigma^2  + h.o.t), 
\;\;\; \sigma \ll 1, \label{J4}
\ee
$\Gamma (\cdot)$ the spatial covariance function of $\xi$, 
$\Gamma (0) > 0$; and:
\be
c^* = O( \sigma ), \;\;\;\sigma \gg 1., \label{J5}
\ee
    
However, if $\xi=\xi(y_1)$, time independent mean zero stationary 
ergodic Gaussian process, 
then almost surely, the front velocity diverges like:
\be
c^{*} \sim c_0 + \sigma \, \sqrt{2 \Gamma(0) \log t }, \;\;\; t\gg 1, 
\label{J6}
\ee
due to the almost sure growth of running 
maxima of Gaussian process. The divergence can be faster for other 
types of noisy processes, for example the Poisson shot noise. 

Comparing (\ref{J4}), 
(\ref{J5}) and (\ref{J6}), we see that the fast switching of 
direction in time helps to slow down fronts, and cures the divergence in 
(\ref{J6}). The slow down of front speeds due to fast direction changing 
in time also occurs in the deterministic setting for all the above 
nonlinearities \cite{NX}. As the front speed asymptotics in the deterministic 
cases (\ref{J2})-(\ref{J3}) are not sensitive to nonlinearities, 
we expect that the analysis on random KPP front speeds, (\ref{J4})-(\ref{J6}),
will have implications for other types of $f(u)$. Numerical work 
along this line is in progress and will be reported elsewhere \cite{NX1}.

The rest of the paper is organized as follows. In section 2, 
we derive the front speed asymptotic formula (\ref{J4}), (\ref{J6}),  
and bound (\ref{J5}), using properties of homogenized 
Hamiltonian and asymptotics of Lyapunov exponents. In section 3, 
we draw conclusions and discuss future works.

\section{KPP Front Speeds in Random Shear Flows}
\setcounter{equation}{0}
The KPP (minimal) front speeds in random media have a 
variational characterization, \cite{Fr1}, \cite{Soug}, \cite{Xin1}, 
and references 
therein, only $f'(0)$ enters from nonlinearity in an otherwise linear problem.

Let $\xi = \xi (y_1,t,\omega )$, a mean zero 
stationary spatial-temporal Gaussian field, with covariance function:
\be
E(\xi(y_1,t)\,\xi(y^{'}_{1},t^{'}))=\Gamma (|t-t'|,|y_1 - y^{'}_{1}|), \label{r0}
\ee
$\forall $ $y$, $y'\in R^1$, $t$, $t' \in R^1$, 
where $\Gamma $ is a positive definite continuous function with decay 
at infinity.  
The KPP front speed formula is:
\be
c^{*} = \inf_{\lam_2 > 0} {H(\lam_1,\lam_2) \over \lam_2}, \label{r1}
\ee
where:
\be
 H(\lam) = f'(0) + \lim_{t \to \infty} \,t^{-1}\, \log \, E [
\exp\{ \lam \cdot Y_{t} \}], \label{r2}
\ee
and $Y_t$ is a diffusion process obeying the Ito equation:
\be
dY_s = d W_s + (0,\sigma \xi(Y_{1,s}, t-s))\, ds,\;\; Y_0 = 0, \label{r3}
\ee
$W_s$ is the standard two dimensional Wiener process. 

\nit It follows from (\ref{r3}) that:
\br
 Y_{1,t} & = & W_{1,t}, \no \\
 dY_{2,s} & =&  dW_{2,s} +\sigma  \xi (W_{1,s},t-s)\, ds. \no
\er
So:
\be
E [ e^{ \lam \cdot Y_{t} }] =
 E [ e^{\lam_1 W_{1,t} + \lam_2\,\sigma  \int_{0}^{t} \xi (W_{1,s},t-s)\, ds } ]
\cdot E [ e^{ \lam_2 W_{2,t}} ].
\label{r4}
\ee
The second factor $= \exp\{ {\lam_{2}^{2}\; t \over  2}\}$. 
The first factor $= u(0,t)$, with $u=u(x,t)$ the solution of:
\br
& &  u_t = {1\over 2} u_{xx} + \lam_2 \, \sigma \xi (x,t)\, u, \;\; x \in R^1, \no \\
& & u(x,0) = e^{\lam_1 x}. \label{r4a}
\er
The initial value problem is invariant in the sense of distribution 
if $(\lam_1,\lam_2) \to -(\lam_1,\lam_2) $, $ x \to -x$, implying that 
the limit: $\lim_{t \to \infty} \, t^{-1}\, \log  u(0,t) \equiv \gamma (\lam)$ 
is an even 
function in $\lam$, if it exists and is nonrandom. 

The function $v = \log u(x,t)$ satisfies 
the viscous Hamilton-Jacobi equation:
\be 
v_{t} = {1\over 2}v_{xx} +{1\over 2}v_{x}^{2} + \lam_2 \, \sigma \xi (x,t),
\label{r5}
\ee
with linear initial data $v(x,0)=\lam_1 \, x$. The limit 
$\lim_{t \to \infty} \, t^{-1}\, v $ agrees with the homogenized 
Hamiltonian of the related homogenization problem, see  
\cite{Xin1} for more exposition of this connection.  
A useful consequence is that $\gamma $ is a convex function in $\lam_1 $ being 
the homogenization limit of a quadratic Hamiltonian with oscillating 
potential (for fixed $\lam_2$), 
see \cite{Rez1}, \cite{Rez2}, \cite{Soug} for analysis of 
convex Hamiltonian and bounded random potentials. 

Evenness and convexity of $\gamma $ in $\lam_1$ implies that
$\gamma (0,\lam_2) \leq \gamma (\lam_1, \lam_2)$, and:
\be
c^{*}
 = \inf_{\lam_2 > 0} {H(0,\lam_2) \over \lam_2}. \label{r6}
\ee

\nit As $\lam_1 = 0$, the problem reduces to the 
{\it parabolic Anderson model},
studied by Carmona, Molchanov, and coworkers, \cite{CM1}, 
\cite{CM2}, \cite{CM3}. The limit, 
\[ \lim_{t \to \infty} \, t^{-1}\, \log  u(0,t) = \gamma, \]
if it exists and is nonrandom, is called the almost sure Lyapunov exponent.

A case where the $\gamma $ limit exists as a finite nonrandom number is 
when $\xi $ is white in time, or $\Gamma (t,y_1) = \dta (t) \Gamma_0(|y_1|)$. 
The spatially discrete case on lattice $Z^d$ was completely 
analyzed in \cite{CM1},
\cite{CM3}, where the Laplacian operator $\Delta_{x}v$ is replaced by 
$\sum_{|x'-x|=1} v(x')-v(x)$.  
The main result is: Let $v$ be the solution of
\be
v_t = \kappa \Delta v + \xi(t,x) \, v, \;\;\; \;\; x \in Z^d, \;\;\; \;\;
 \kappa > 0, \label{r7}
\ee
with product in the Stratonovich sense, then the almost sure 
Lyapunov exponent $\gamma (\kappa)$ exists as 
nonrandom number and obeys 
($b_0 > 0$, $d=1, 2$):
\br
\gamma (\kappa ) & \sim & {b_0 \over -\log \kappa}, \;\;\;\;\; \kappa \ll 1, 
\label{r8a} \\
 \gamma (\kappa ) & \sim & {\Gamma (0)\over 2}, \;\;\;\;\;
 \kappa \gg 1. \label{r8b}
\er
In three and higher dimensions, $\gamma (\kappa ) = \Gamma (0)/2$ if 
$\kappa $ exceeds a critical level $\kappa_{cr} $. This shows 
that the larger the diffusion effect, the more $\gamma $ tends to be  
constant $\Gamma (0)/2$.   Moreover, $\gamma(\kappa )$ is 
a monotone increasing continuous function in $\kappa$ \cite{CM1}. 
The continuum case is 
more involved, Cranston and Montford just proved \cite{Cr} that 
$\gamma(\kappa) \sim c_1 \kappa^{p}$, $p\in (0,1/2)$, $c_1 > 0$, 
$\kappa \ll 1$, and that $\gamma (\kappa) > 0$ for all $\kappa > 0$. The 
large $\kappa $ regime (\ref{r8b}) is expected to be the same, also 
$\gamma (\kappa )$ is monotone in $\kappa $. If 
the continuum $\Delta $ is approximated by standard central differencing, 
we see that the continuum $\Delta $ is close to a discrete one with 
larger diffusion.

\nit The results can be adapted to equation (\ref{r4a}) with $\lam_1 =0$. 
First ${1\over \sqrt{a}}\xi(t/a,y) = \xi(t,y)$ in law, for any $a > 0$. 
Let $t = t'/a$, $a = (\sigma \lam_{2})^{2}$, equation (\ref{r4a}) becomes:
\be
u_{t'} = {1 \over 2 \sigma^{2} \lam_{2}^{2}} \Delta \, u + \xi(t',x)\, u, 
\label{r9}
\ee
It follows that $\lim_{t \to \infty} u(t,0)/t = \gamma^{*}(\sigma \lam_2)$ 
such that
\br
\gamma^{*}(\sigma \lam_2) & \sim & {\Gamma (0)\, (\sigma \lam_{2})^{2} \over 2}, \;\; 
\;\;\; \sigma \lam_{2} \ll 1, \label{r10} \\
\gamma^{*}(\sigma \lam_2) & \sim & b_0 h
\left ({1\over 2 \sigma^2 \lam_{2}^{2}}\right )\, (\sigma \lam_2)^2, 
\;\;\;\;\sigma \lam_{2} \gg 1, \label{r11}
\er
where $h=h(x) = x^{p}$, $p \in (0,1/2)$.
Now we minimize $H(0,\lam_2)/\lam_2 $. 

For $\sigma \ll 1$, and $\lam_2 \sim O(1)$, where the minimizer is expected:  
\[H(0,\lam_2)/\lam_2 \sim f'(0)/\lam_2 + \lam_2/2 + 
{\Gamma (0)\, \sigma^{2} \lam_{2} \over 2} + h.o.t., \]
giving minimal point:
\[ \lam_{2} \sim \sqrt{2 f'(0)\over 1 + \Gamma (0)\sigma^2}, \]
and minimum:
\be
c^{*} = \sqrt{2 f'(0)}\, (1 + {1\over 2} \Gamma (0)\sigma^2 ) + h.o.t.  \label{r12}
\ee
This is the stochastic analogue of quadratic speed enhancement. 

For $\sigma \gg 1$, and $\lam_2 \sigma \gg 1$, we have:
\[  H(0,\lam_2)/\lam_2 \sim b_0 h\left ({1\over  2 \sigma^2 \lam_{2}^{2}}\right ) 
\sigma^2 \lam_{2} + f'(0)/\lam_2 + \lam_2/2, \]
giving an upper bound with $\lam_2 = {\sigma_0 \over \sigma}$, $\sigma_0$ 
large enough constant so that $\sigma \gg \sigma_0 \gg 1$:
\be
c^{*} = \inf_{\lam_2 > 0} H(0,\lam_2)/\lam_2 \leq O( \sigma_{0}^{1-2p} 
\sigma), 
\label{r13}
\ee
which shows that the upper bound on the speed is linear in $\sigma \gg 1$. 

To establish a similar lower bound, let us note that
$\gamma (\kappa )$ is bounded away from zero 
by an amount $\gamma_0 $ if $\kappa \geq \kappa_0$, and asymptotics 
$\gamma(\kappa) \sim c_1 \kappa^{p}$ holds, for some $p\in (0,1/2)$, 
if $\kappa < \kappa_0$. 

Now we estimate $c^{*}$ from below. 
If $2(\sigma \lam_2)^{2} \leq \kappa_{0}^{-1}$, then:
\be
H(0,\lam_2)/\lam_2 \geq \gamma_0 \sigma^2 \lam_2 + f'(0)/\lam_2 + 
\lam_2/2 \equiv F(\lam_2). \label{r14}
\ee
The absolute minimal point $\lam_{2,m}$ of $F$ satisfies:
\[ 1/2 -f'(0)/\lam_{2}^{2} +\gamma_{0}\sigma^{2} = 0, \]
or:
\be
 \lam_{2,m} = \sqrt{ {2f'(0)\over 1 + 2\gamma_0 \sigma^2}} \sim 
O(\sigma^{-1}), \;\;\;\;\sigma \gg 1. \label{r15}
\ee 
So $\lam_{2,m}\sigma = O(1)$, and $
H(0,\lam_2)/\lam_2 \geq F(\lam_2) \geq O(\sigma )$.

On the other hand, if $2(\sigma \lam_2)^{2} \geq \kappa_{0}^{-1}$,
\be
H(0,\lam_2)/\lam_2 \geq c_1 2^{-p} \sigma^{2 - 2p}\lam_{2}^{1 -2p} 
+ f'(0)/\lam_2 +\lam_2/2 \equiv G(\lam_2). \label{r16}
\ee
The absolute minimal point $\lam_{2,M}$ of $G$ obeys:
\[ c_1 2^{-p}(1-2p)\sigma^{2 -2p}\lam_{2}^{-2p} + 1/2  = f'(0)/\lam_{2}^{2},
\]
implying:
\[ c_1 2^{-p}(1-2p)\sigma^{2 -2p}\lam_{2,M}^{-2p} \leq f'(0)/\lam_{2,M}^{2},\]
or:
\be
 \lam_{2,M} \leq C(p,c_1,f'(0))/ \sigma, \label{r17}
\ee
for some positive constant $C(p,c_1,f'(0))$. It follows from 
(\ref{r17}) and (\ref{r16}) that:
\[ H(0,\lam_2)/\lam_2 \geq f'(0)/\lam_{2,M} \geq C'(p,c_1,f'(0))\sigma. \]
The above two lower bounds combine to give $c^{*} \geq O(\sigma )$, 
and together with upper bound (\ref{r13}), we have the front speed 
bound $c^{*} = O(\sigma )$ for large $\sigma $.

The $\gamma $ diverges logarithmically 
when $\xi =\xi(x)$ is a spatial 
Gaussian process \cite{CM3}. The leading order exponential 
growth rate of $u$ depends on the fact that:
\be
\sup_{|x| \leq t}\, \xi (x) \sim \sqrt{2\Gamma (0) \, \log t},  \;\;\; 
t \to \infty, \;\; a.s.,\label{r14a}
\ee
and so:
\be
\lim_{t \to \infty}\, {1\over t\sqrt{\log t}}\, \log\, u(t,x) = 
\sigma |\lam_2| \sqrt{2\Gamma (0) }, \;\; a.s.\label{r15a}
\ee
implying front speed divergence in time. In other words, at large $t$:
\[
{H(0,\lam_2) \over \lam_2} \sim {f'(0)\over \lam_2} 
+{\lam_2 \over 2} + \sigma sign(\lam_2) \sqrt{2\Gamma (0)\log t}, \]
or:
\be
 c^{*} = \inf_{\lam_2 > 0} {H(0,\lam_2)\over \lam_2} 
\sim c_0 + \sigma \sqrt{2\Gamma (0)\log t}.   \label{r16a}
\ee

In view of (\ref{r14a}), such divergence can be avoided 
if the front is restricted inside an infinite  
channel of finite cross section. 
However, the almost sure constant speed ceases to exist, 
as on a finite interval $\xi$ process loses ergodicity, 
different realizations can be either more positive or negative, so  
ensemble averaged speed becomes an alternative measure of front speed.
A numerical study of random fronts in channels will be reported 
elsewhere \cite{NX1}. 

\section{Conclusions}
We found that KPP front speeds are finite and enhanced through 
 white in time stationary Gaussian random shears. The enhancement is 
quadratic in the root mean square (rms) of small shear amplitudes, and grows 
at the order of  the large rms amplitudes. The front speeds diverge 
logarithmically in time independent mean zero Gaussian shears. 
The findings will help to investigate similar front behaviors 
under other forms of nonlinearities, and serve as a useful guide for 
numerical works on fronts through random shears in channels of 
large but finite cross sections. 

\section{Acknowledgements}
The work was partially done during my visit to Inst H. Poincar\'e (IHP) in 
Oct, 2002. I would like to thank H. Berestycki and J-M Roquejoffre 
for inviting and their hospitalities, as well as IHP for a visiting 
professorship. 
I also thank S. Molchanov and M. Cranston for communicating 
ongoing work on parabolic Anderson models. Partial NSF support is 
gratefully acknowledged.

\bibliographystyle{plain}

\begin{thebibliography}{99}

\bibitem{B1} B. Audoly, H. Berestycki, Y. Pomeau,
{\em Raction-Diffusion en e\'coulement rapide}, 
Note C. R. Acad. Sc. Paris 328, S\'erie II, 2000, pp. 255-262.

\bibitem{Surf}A-L Barab\'asi, H. E. Stanley,
{\em Fractal Concepts in Surface Growth}, Cambridge University Press, 1995. 

\bibitem{B2}H. Berestycki, 
{\em The influence of advection on the propagation of fronts 
in reaction-diffusion equations}, Proceedings NATO ASI Conf. Cargese, 
H. Berestycki and Y. Pomeau eds, Kluwer, to appear. 

\bibitem{CM1}R. Carmona, S. Molchanov,
{\em Parabolic Anderson Model and Intermittency},
Memoirs of the AMS, 108, No. 518 (1994).

\bibitem{CM2}R. Carmona, S. Molchanov,
{\em Stationary parabolic Anderson model and intermittency},
Prob. Theory Rel. Fields 102, 433-453 (1995).

\bibitem{CM3}R. Carmona, L. Koralov, S. Molchanov,
{\em Asymptotics for the almost sure Lyapunov exponent for the
solution of the parabolic Anderson problem}, Random Oper. and Stoch. Equ.,
Vol. 9, No. 1, pp. 77-86 (2001).

\bibitem{Const1} P. Constantin, A. Kiselev, A. Oberman, L. Ryzhik, 
{\em Bulk burning rate in passive-reactive diffusion}, Arch Rat. 
Mech Analy, 154, 2000, 53-91. 

\bibitem{Cr}M. Cranston,
{\em personal communication}, 2002.

\bibitem{Fr1}M. Freidlin,
{\em Functional Integration and Partial Differential Equations.}
Annals of Mathematics Studies. Number 109. Princeton University Press, 1985.

\bibitem{H}S. Heinze,
{\em The speed of travelling waves for convective reaction-diffusion 
equations}, preprint 84, 
Max-Planck-Institut f\"ur Mathematik in den Naturewissenschaften, 
Leipzig, 2001.
 
\bibitem{HPS}S. Heinze, G. Papanicolaou, A. Stevens, 
{\em Variational principles for propagation speeds in inhomogeneous media}, 
SIAM J. Applied Math, 62, no. 1, 2001.

\bibitem{KR}A. Kiselev, L. Ryzhik,
{\em Enhancement of the traveling front speeds in reaction-diffusion 
equations with advection}, Ann. de l'Inst. Henri Poincar\'e, 
Analyse Nonlin\'eaire, 18, 2001, 309--358.

\bibitem{Molch}S. Molchanov,
{\em personal communication}, 2002. 

\bibitem{NX}J. Nolen, J. Xin,
{\em Analysis and computation of reaction-diffusion front speeds 
in spatially-temporally periodic 
shear flows}, preprint, 2002.

\bibitem{NX1} J. Nolen, J. Xin,
{\em A numerical study of reaction-diffusion front speeds in 
random shear flows}, in preparation.

\bibitem{PX}G. Papanicolaou, J. Xin,
{\em Reaction-Diffusion Fronts in Periodically Layered Media}, 
J. Stat. Physics 63(1991), pp 915-931.

\bibitem{Rez1}F. Rezakhanlou,
{\em Central Limit Theorem for Stochastic Hamilton-Jacobi Equations},
Comm. Math Physics, 211(2000), pp 413-438.

\bibitem{Rez2}F. Rezakhanlou, J. Tarver,
{\em Homogenization for Stochastic Hamilton-Jacobi Equations},
Arch. Rat. Mech. Anal. 151(2000), pp 277-309.

\bibitem{Soug}P. E. Souganidis,
{\em Stochastic homogenization of Hamilton-Jacobi equations and some
applications}, Asymptotic Analysis 20 (1999), pp 1-11.

\bibitem{Const2}N. Vladimirova, P. Constantin, A. Kiselev, O. Ruchaiskiy, 
L. Ryzhik,
{\em Flame Enhancement and Quenching in Fluid Flows}, preprint, 2002. 

\bibitem{Xin1} J. Xin,
{\em Front Propagation in Heterogeneous Media,}
SIAM Review, Vol. 42, No. 2, June 2000, pp 161-230. 


\end{thebibliography}

\end{document}